\documentclass[12pt,a4paper]{article}

\usepackage{ifthen} 
\newboolean{pdflatex}
\setboolean{pdflatex}{true} 

\newboolean{articletitles}
\setboolean{articletitles}{true} 

\newboolean{uprightparticles}
\setboolean{uprightparticles}{false} 

\newboolean{inbibliography}
\setboolean{inbibliography}{false} 

\usepackage{microtype}
\usepackage{lineno}  
\usepackage{xspace} 

\usepackage{graphicx}  
\usepackage{color}
\usepackage{colortbl}
\graphicspath{{.}{./figs/}} 

\usepackage{amsmath} 
\usepackage{amssymb}
\usepackage{amsfonts}
\usepackage{upgreek} 

\newcommand*\patchAmsMathEnvironmentForLineno[1]{%
\expandafter\let\csname old#1\expandafter\endcsname\csname #1\endcsname
\expandafter\let\csname oldend#1\expandafter\endcsname\csname
end#1\endcsname
 \renewenvironment{#1}%
   {\linenomath\csname old#1\endcsname}%
   {\csname oldend#1\endcsname\endlinenomath}%
}
\newcommand*\patchBothAmsMathEnvironmentsForLineno[1]{%
  \patchAmsMathEnvironmentForLineno{#1}%
  \patchAmsMathEnvironmentForLineno{#1*}%
}
\AtBeginDocument{%
\patchBothAmsMathEnvironmentsForLineno{equation}%
\patchBothAmsMathEnvironmentsForLineno{align}%
\patchBothAmsMathEnvironmentsForLineno{flalign}%
\patchBothAmsMathEnvironmentsForLineno{alignat}%
\patchBothAmsMathEnvironmentsForLineno{gather}%
\patchBothAmsMathEnvironmentsForLineno{multline}%
}

\usepackage{hyperref}    
\usepackage[all]{hypcap} 

\def\lhcb {\mbox{LHCb}\xspace}

\def\babar  {\mbox{BaBar}\xspace}
\def\belle  {\mbox{Belle}\xspace}

\def\velo   {VELO\xspace}

\ifthenelse{\boolean{uprightparticles}}%
{

 \def\Ppi         {\ensuremath{\uppi}\xspace}

 \def\Pphi        {\ensuremath{\upphi}\xspace}

 \def\Ppsi        {\ensuremath{\uppsi}\xspace}

 \def\PDelta      {\ensuremath{\Delta}\xspace}                 
 \def\PXi      {\ensuremath{\Xi}\xspace}                 
 \def\PLambda      {\ensuremath{\Lambda}\xspace}                 
 \def\PSigma      {\ensuremath{\Sigma}\xspace}                 
 \def\POmega      {\ensuremath{\Omega}\xspace}                 
 \def\PUpsilon      {\ensuremath{\Upsilon}\xspace}                 
 
 \def\PB      {\ensuremath{\mathrm{B}}\xspace}                 
                  
 \def\PD      {\ensuremath{\mathrm{D}}\xspace}

 \def\PJ      {\ensuremath{\mathrm{J}}\xspace}                 
 \def\PK      {\ensuremath{\mathrm{K}}\xspace}

 \def\PX      {\ensuremath{\mathrm{X}}\xspace}

 \def\Pb      {\ensuremath{\mathrm{b}}\xspace}                 
 \def\Pc      {\ensuremath{\mathrm{c}}\xspace}

 \def\Pi      {\ensuremath{\mathrm{i}}\xspace}

 \def\Pp      {\ensuremath{\mathrm{p}}\xspace}

 \def\Ps      {\ensuremath{\mathrm{s}}\xspace}

}
{

 \def\Ppi         {\ensuremath{\pi}\xspace}

 \def\Pphi        {\ensuremath{\phi}\xspace}

 \def\Ppsi        {\ensuremath{\psi}\xspace}                 
                  
 \mathchardef\PDelta="7101
 \mathchardef\PXi="7104
 \mathchardef\PLambda="7103
 \mathchardef\PSigma="7106
 \mathchardef\POmega="710A
 \mathchardef\PUpsilon="7107
                  
 \def\PB      {\ensuremath{B}\xspace}                 
                  
 \def\PD      {\ensuremath{D}\xspace}

 \def\PJ      {\ensuremath{J}\xspace}                 
 \def\PK      {\ensuremath{K}\xspace}

 \def\PX      {\ensuremath{X}\xspace}

 \def\Pb      {\ensuremath{b}\xspace}                 
 \def\Pc      {\ensuremath{c}\xspace}

 \def\Pi      {\ensuremath{i}\xspace}

 \def\Pp      {\ensuremath{p}\xspace}

 \def\Ps      {\ensuremath{s}\xspace}

}

\def\squark    {\ensuremath{\Ps}\xspace}

\def\cquark    {\ensuremath{\Pc}\xspace}

\def\bquark    {\ensuremath{\Pb}\xspace}
\def\bquarkbar {\ensuremath{\overline \bquark}\xspace}

\def\pion  {\ensuremath{\Ppi}\xspace}

\def\pip   {\ensuremath{\pion^+}\xspace}
\def\pim   {\ensuremath{\pion^-}\xspace}
\def\pipm  {\ensuremath{\pion^\pm}\xspace}

\def\kaon  {\ensuremath{\PK}\xspace}
  \def\Kbar  {\kern 0.2em\overline{\kern -0.2em \PK}{}\xspace}

\def\Kz    {\ensuremath{\kaon^0}\xspace}
\def\Kzb   {\ensuremath{\Kbar^0}\xspace}
\def\Kp    {\ensuremath{\kaon^+}\xspace}
\def\Km    {\ensuremath{\kaon^-}\xspace}
\def\Kpm   {\ensuremath{\kaon^\pm}\xspace}

\def\KS    {\ensuremath{\kaon^0_{\rm\scriptscriptstyle S}}\xspace} 
\def\KL    {\ensuremath{\kaon^0_{\rm\scriptscriptstyle L}}\xspace} 

\def\Kstarzb {\ensuremath{\Kbar^{*0}}\xspace}

  \def\Dbar    {\kern 0.2em\overline{\kern -0.2em \PD}{}\xspace}
\def\D       {\ensuremath{\PD}\xspace}

\def\Dz      {\ensuremath{\D^0}\xspace}

\def\Dstarp  {\ensuremath{\D^{*+}}\xspace}

\def\B       {\ensuremath{\PB}\xspace}
\def\Bbar    {\ensuremath{\kern 0.18em\overline{\kern -0.18em \PB}{}}\xspace}

\def\Bz      {\ensuremath{\B^0}\xspace}

\def\Bu      {\ensuremath{\B^+}\xspace}
\def\Bub     {\ensuremath{\B^-}\xspace}
\def\Bp      {\ensuremath{\Bu}\xspace}
\def\Bm      {\ensuremath{\Bub}\xspace}
\def\Bpm     {\ensuremath{\B^\pm}\xspace}

\def\Bs      {\ensuremath{\B^0_\squark}\xspace}

\def\Bc      {\ensuremath{\B_\cquark^+}\xspace}
\def\Bcp     {\ensuremath{\B_\cquark^+}\xspace}

\def\jpsi     {\ensuremath{{\PJ\mskip -3mu/\mskip -2mu\Ppsi\mskip 2mu}}\xspace}

  \def\Y#1S{\ensuremath{\PUpsilon{(#1S)}}\xspace}

\def\proton      {\ensuremath{\Pp}\xspace}

\def\L {\ensuremath{\PLambda}\xspace}
\def\Lbar {\ensuremath{\kern 0.1em\overline{\kern -0.1em\PLambda}}\xspace}

\def\Lb      {\ensuremath{\L^0_\bquark}\xspace}

\def\BF         {{\ensuremath{\cal B}\xspace}}

\newcommand{\decay}[2]{\ensuremath{#1\!\to #2}\xspace}         

\def\to                 {\ensuremath{\rightarrow}\xspace}

\def\eps   {\ensuremath{\varepsilon}\xspace}

\def\CP                {\ensuremath{C\!P}\xspace}

\newcommand{\ACP}{\ensuremath{{\cal A}^{\CP}}\xspace}

\def\AT#1     {\ensuremath{A_{\mathrm{T}}^{#1}}\xspace}           

\def\C#1      {\ensuremath{\mathcal{C}_{#1}}\xspace}                       
\def\Cp#1     {\ensuremath{\mathcal{C}_{#1}^{'}}\xspace}                    
\def\Ceff#1   {\ensuremath{\mathcal{C}_{#1}^{\mathrm{(eff)}}}\xspace}        
\def\Cpeff#1  {\ensuremath{\mathcal{C}_{#1}^{'\mathrm{(eff)}}}\xspace}       
\def\Ope#1    {\ensuremath{\mathcal{O}_{#1}}\xspace}                       
\def\Opep#1   {\ensuremath{\mathcal{O}_{#1}^{'}}\xspace}                    

\newcommand{\tev}{\ifthenelse{\boolean{inbibliography}}{\ensuremath{~T\kern -0.05em eV}\xspace}{\ensuremath{\mathrm{\,Te\kern -0.1em V}}\xspace}}
\newcommand{\gev}{\ensuremath{\mathrm{\,Ge\kern -0.1em V}}\xspace}
\newcommand{\mev}{\ensuremath{\mathrm{\,Me\kern -0.1em V}}\xspace}
\newcommand{\kev}{\ensuremath{\mathrm{\,ke\kern -0.1em V}}\xspace}
\newcommand{\ev}{\ensuremath{\mathrm{\,e\kern -0.1em V}}\xspace}
\newcommand{\gevc}{\ensuremath{{\mathrm{\,Ge\kern -0.1em V\!/}c}}\xspace}
\newcommand{\mevc}{\ensuremath{{\mathrm{\,Me\kern -0.1em V\!/}c}}\xspace}
\newcommand{\gevcc}{\ensuremath{{\mathrm{\,Ge\kern -0.1em V\!/}c^2}}\xspace}
\newcommand{\gevgevcccc}{\ensuremath{{\mathrm{\,Ge\kern -0.1em V^2\!/}c^4}}\xspace}
\newcommand{\mevcc}{\ensuremath{{\mathrm{\,Me\kern -0.1em V\!/}c^2}}\xspace}

\def\mm   {\ensuremath{\rm \,mm}\xspace}

\def\mum  {\ensuremath{\,\upmu\rm m}\xspace}

\def\invfb   {\ensuremath{\mbox{\,fb}^{-1}}\xspace}

\newcommand{\chisq}{\ensuremath{\chi^2}\xspace}

\newcommand{\chisqip}{\ensuremath{\chi^2_{\rm IP}}\xspace}
\newcommand{\chisqvs}{\ensuremath{\chi^2_{\rm VS}}\xspace}

\def\gsim{{~\raise.15em\hbox{$>$}\kern-.85em
          \lower.35em\hbox{$\sim$}~}\xspace}
\def\lsim{{~\raise.15em\hbox{$<$}\kern-.85em
          \lower.35em\hbox{$\sim$}~}\xspace}

\def\ptot       {\mbox{$p$}\xspace}
\def\pt         {\mbox{$p_{\rm T}$}\xspace}

\def\dllkpi     {\ensuremath{\mathrm{DLL}_{\kaon\pion}}\xspace}

\def\dllpk      {\ensuremath{\mathrm{DLL}_{\proton\kaon}}\xspace}

\def\mrad{\ensuremath{\rm \,mrad}\xspace}

\def\evtgen     {\mbox{\textsc{EvtGen}}\xspace}

\def\geant      {\mbox{\textsc{Geant4}}\xspace}

\def\photos     {\mbox{\textsc{Photos}}\xspace}

\def\pythia     {\mbox{\textsc{Pythia}}\xspace}

\def\tell1  {TELL1\xspace}
\def\ukl1   {UKL1\xspace}

\ifthenelse{\boolean{uprightparticles}}%
{
 \def\Pphi      {\ensuremath{\upphi}\xspace}
}
{
 \def\Pphi      {\ensuremath{\phi}\xspace}
}

\def\kpi       {\ensuremath{\kaon\pion}\xspace}

\def\KSK   {\ensuremath{\KS\Kpm}\xspace}
\def\KSPi   {\ensuremath{\KS\pipm}\xspace}
\def\KSKp   {\ensuremath{\KS\Kp}\xspace}
\def\KSPip   {\ensuremath{\KS\pip}\xspace}

\def\ButoKsK   {\decay{\Bp}{\KS \Kp}}
\def\ButoKsPi   {\decay{\Bp}{\KS \pip}}
\def\BptoKSK   {\decay{\Bp}{\KS \Kp}}
\def\BptoKSPi   {\decay{\Bp}{\KS \pip}}
\def\BmtoKSK   {\decay{\Bm}{\KS \Km}}
\def\BmtoKSPi   {\decay{\Bm}{\KS \pim}}

\def\BctoKsK   {\decay{\Bcp}{\KS \Kp}}

\def\BctoKzK   {\decay{\Bcp}{\Kzb \Kp}}
\def\BctoKstK   {\decay{\Bcp}{\Kstarzb \Kp}}
\def\BctoPhiK   {\decay{\Bcp}{\Pphi \Kp}}
\def\ButoKzPi   {\decay{\Bp}{\Kz \pip}}

\def\ButoJpsiK   {\decay{\Bpm}{\jpsi \Kpm}}

\newcommand{\Br}[1]{\ensuremath{\BF\left(#1\right)}\xspace}
\newcommand{\Acp}[1]{\ensuremath{\ACP\left(#1\right)}\xspace}
\newcommand{\Gam}[1]{\ensuremath{\Gamma\left(#1\right)}\xspace}

\usepackage{cite} 
\usepackage{mciteplus}

\usepackage{longtable} 

\begin{document}

\renewcommand{\thefootnote}{\fnsymbol{footnote}}
\setcounter{footnote}{1}


\begin{titlepage}
\pagenumbering{roman}

\vspace*{-1.5cm}
\centerline{\large EUROPEAN ORGANIZATION FOR NUCLEAR RESEARCH (CERN)}
\vspace*{1.5cm}
\hspace*{-0.5cm}
\begin{tabular*}{\linewidth}{lc@{\extracolsep{\fill}}r}
\ifthenelse{\boolean{pdflatex}}
{\vspace*{-2.7cm}\mbox{\!\!\!\includegraphics[width=.14\textwidth]{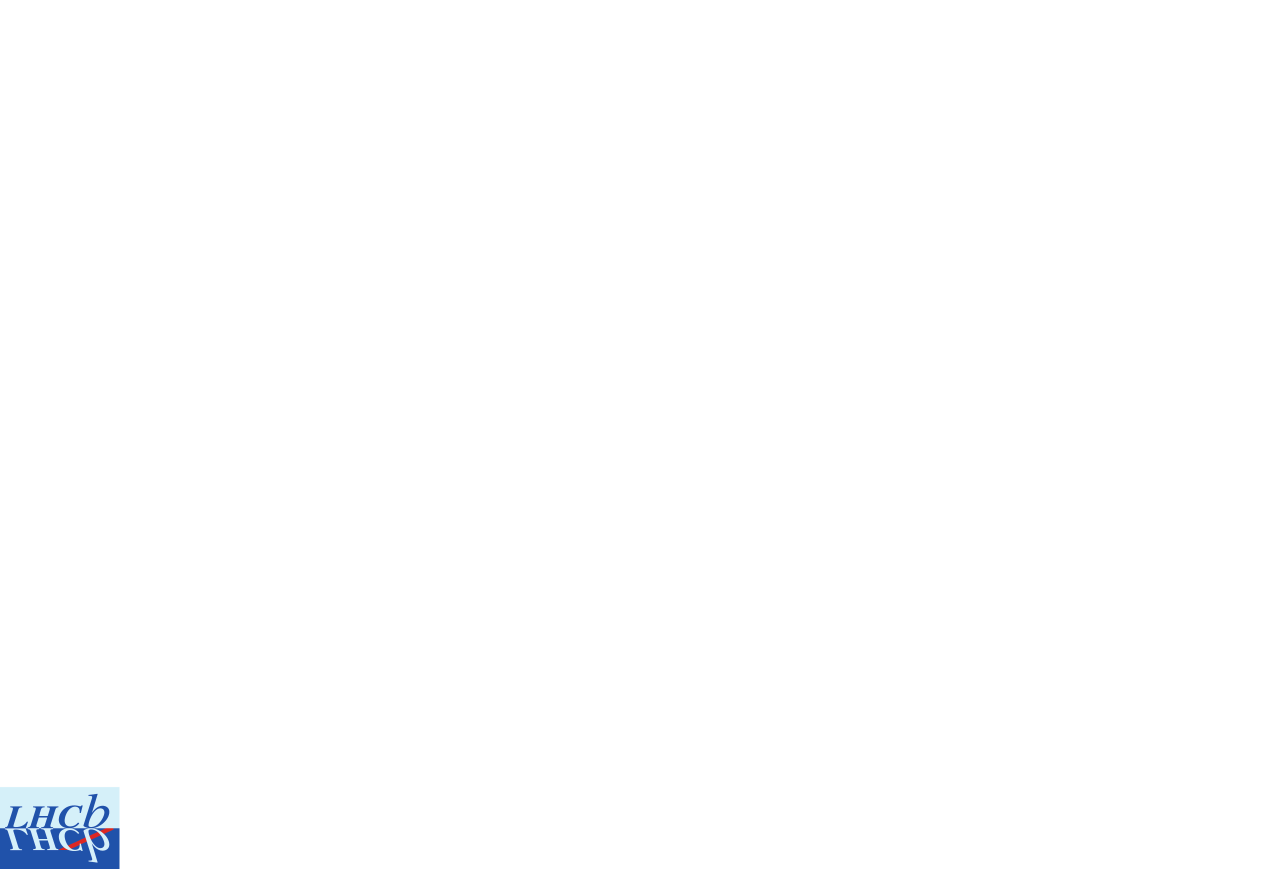}} & &}%
{\vspace*{-1.2cm}\mbox{\!\!\!\includegraphics[width=.12\textwidth]{lhcb-logo.eps}} & &}%
\\
 & & CERN-PH-EP-2013-147 \\  
 & & LHCb-PAPER-2013-034 \\  
 & & October 8, 2013 \\ 
 & & \\
\end{tabular*}

\vspace*{2.0cm}

{\bf\boldmath\huge
\begin{center}
  Branching fraction and \textbf{\textit{CP}} asymmetry of the decays $B^+ \to K_{\rm \scriptscriptstyle S}^0 \pi^+$ and $B^+ \to K_{\rm \scriptscriptstyle S}^0 K^+$
\end{center}
}

\vspace*{1.0cm}

\begin{center}
The LHCb collaboration\footnote{Authors are listed on the following pages.}
\end{center}

\vspace*{1.0cm}

\begin{abstract}
  \noindent
 An analysis of $B^+ \to K_{\rm \scriptscriptstyle S}^0 \pi^+$ and $B^+ \to K_{\rm \scriptscriptstyle S}^0 K^+$ decays is performed with the LHCb experiment. The $pp$ collision data used correspond to integrated luminosities of $1\mbox{\,fb}^{-1}$ and $2\mbox{\,fb}^{-1}$ collected at centre-of-mass energies of $\sqrt{s}=7\mathrm{\,Te\kern -0.1em V}$ and $\sqrt{s}=8\mathrm{\,Te\kern -0.1em V}$, respectively. The ratio of branching fractions and the direct {\it CP} asymmetries are measured to be $\mathcal{B}(B^+ \to K_{\rm \scriptscriptstyle S}^0 K^+)/\mathcal{B}(B^+ \to K_{\rm \scriptscriptstyle S}^0 \pi^+) =  0.064 \pm 0.009\textrm{ (stat.)} \pm 0.004\textrm{ (syst.)}$, $\mathcal{A}^{\it CP}(B^+ \to K_{\rm \scriptscriptstyle S}^0 \pi^+) = -0.022 \pm 0.025\textrm{ (stat.)}  \pm 0.010\textrm{ (syst.)}$ and $\mathcal{A}^{\it CP}(B^+ \to K_{\rm \scriptscriptstyle S}^0 K^+) = -0.21 \pm 0.14\textrm{ (stat.)}  \pm 0.01\textrm{ (syst.)}$. The data sample taken at $\sqrt{s}=7\mathrm{\,Te\kern -0.1em V}$ is used to search for $B_c^+ \to K_{\rm \scriptscriptstyle S}^0 K^+$ decays and results in the upper limit $(f_c\cdot\mathcal{B}(B_c^+ \to K_{\rm \scriptscriptstyle S}^0 K^+))/(f_u\cdot\mathcal{B}(B^+ \to K_{\rm \scriptscriptstyle S}^0 \pi^+)) < 5.8\times10^{-2}\textrm{  at 90\% confidence level}$, where $f_c$ and $f_u$ denote the hadronisation fractions of a $\bar{b}$  quark into a $B_c^+$ or a $B^+$ meson, respectively.
\end{abstract}

\vspace*{1.0cm}

\begin{center}
  Submitted to Phys.~Lett.~B
\end{center}

\vspace{\fill}

{\footnotesize 
\centerline{\copyright~CERN on behalf of the \lhcb collaboration, license \href{http://creativecommons.org/licenses/by/3.0/}{CC-BY-3.0}.}}
\vspace*{2mm}

\end{titlepage}


\newpage
\setcounter{page}{2}
\mbox{~}
\newpage


\centerline{\large\bf LHCb collaboration}
\begin{flushleft}
\small
R.~Aaij$^{40}$, 
B.~Adeva$^{36}$, 
M.~Adinolfi$^{45}$, 
C.~Adrover$^{6}$, 
A.~Affolder$^{51}$, 
Z.~Ajaltouni$^{5}$, 
J.~Albrecht$^{9}$, 
F.~Alessio$^{37}$, 
M.~Alexander$^{50}$, 
S.~Ali$^{40}$
% ...

{\footnotesize \it
$ ^{1}$Centro Brasileiro de Pesquisas F\'{i}sicas (CBPF), Rio de Janeiro, Brazil\\
% ...
$ ^{5}$Clermont Universit\'{e}, Universit\'{e} Blaise Pascal, CNRS/IN2P3, LPC, Clermont-Ferrand, France\\
$ ^{6}$CPPM, Aix-Marseille Universit\'{e}, CNRS/IN2P3, Marseille, France\\
% ...
$ ^{9}$Fakult\"{a}t Physik, Technische Universit\"{a}t Dortmund, Dortmund, Germany\\
% ...
$ ^{36}$Universidad de Santiago de Compostela, Santiago de Compostela, Spain\\
$ ^{37}$European Organization for Nuclear Research (CERN), Geneva, Switzerland\\
% ...
$ ^{40}$Nikhef National Institute for Subatomic Physics, Amsterdam, The Netherlands\\
% ...
$ ^{45}$H.H. Wills Physics Laboratory, University of Bristol, Bristol, United Kingdom\\
% ...
$ ^{50}$School of Physics and Astronomy, University of Glasgow, Glasgow, United Kingdom\\
$ ^{51}$Oliver Lodge Laboratory, University of Liverpool, Liverpool, United Kingdom\\
% ...
$ ^{b}$Universit\`{a} di Bari, Bari, Italy\\
% ...
$ ^{s}$Scuola Normale Superiore, Pisa, Italy\\
}
\end{flushleft}
 

\cleardoublepage


\renewcommand{\thefootnote}{\arabic{footnote}}
\setcounter{footnote}{0}



\pagestyle{plain} 
\setcounter{page}{1}
\pagenumbering{arabic}


\section{Introduction}
\label{sec:Introduction}

Studies of charmless two-body \B meson decays allow tests of the Cabibbo-Kobayashi-Maskawa picture of \CP violation \cite{Cabibbo:1963yz,Kobayashi:1973fv} in the Standard Model (SM).
They include contributions from loop amplitudes, and are therefore particularly sensitive to processes beyond the SM \cite{Fleischer:1999pa,Gronau:2000md,Lipkin:2005pb,Fleischer:2007hj,Fleischer:2010ib}.
However, due to the presence of poorly known hadronic parameters, predictions of \CP violating asymmetries and branching fractions are imprecise.
This limitation may be overcome by combining measurements from several charmless two-body \B meson decays and using flavour symmetries~\cite{Fleischer:1999pa}.
More precise measurements of the branching fractions and \CP violating asymmetries will improve the determination of the size of SU(3) breaking effects and the magnitudes of colour-suppressed and annihilation amplitudes~\cite{Buras:2003dj,Baek:2007yy}.

In \ButoKsK and \ButoKsPi decays,\footnote{The inclusion of charge conjugated decay modes is implied throughout this Letter unless otherwise stated.} gluonic loop, colour-suppressed electroweak loop and annihilation amplitudes contribute. Measurements of their branching fractions and \CP asymmetries allow to check for the presence of sizeable contributions from the latter two~\cite{Fleischer:2007hj}. Further flavour symmetry checks can also be performed by studying these decays \cite{He:2013vta}.
First measurements have been performed by the \babar and \belle experiments \cite{Aubert:2006gm,Duh:2012ie}. The world averages are $\Acp{\ButoKsPi} = -0.015 \pm 0.019$, $\Acp{\ButoKsK} = 0.04 \pm 0.14$ and $\Br{\ButoKsK}/\Br{\ButoKsPi} = 0.050 \pm 0.008$, where 
\begin{eqnarray}
\Acp{\ButoKsPi}  & \equiv & \frac{\Gam{\BmtoKSPi}-\Gam{\ButoKsPi}}{{\Gam{\BmtoKSPi}+\Gam{\ButoKsPi}}}
\end{eqnarray}
and $\Acp{\ButoKsK}$ is defined in an analogous way.

Since the annihilation amplitudes are expected to be small in the SM and are often accompanied by other topologies, they are difficult to determine unambiguously. These can however be measured cleanly in \BctoKsK decays, where other amplitudes do not contribute. Standard Model predictions for the branching fractions of pure annihilation \Bcp decays range from $10^{-8}$ to $10^{-6}$ depending on the theoretical approach employed~\cite{PhysRevD.80.114031}.

In this Letter, a measurement of the ratio of branching fractions of \ButoKsK and \ButoKsPi decays with the \lhcb detector is reported along with a determination of their \CP asymmetries. The data sample corresponds to integrated luminosities of 1 and 2\invfb, recorded during 2011 and 2012 at centre-of-mass energies of 7 and 8\tev, respectively. A search for the pure annihilation decay \BctoKsK based on the data collected at 7\tev is also presented. The \ButoKsK and \BctoKsK signal regions, along with the raw \CP asymmetries, were not examined until the event selection and the fit procedure were finalised.

\section{Detector, data sample and event selection}
\label{sec:Selection}

The \lhcb detector~\cite{Alves:2008zz} is a single-arm forward
spectrometer covering the \mbox{pseudorapidity} range $2<\eta <5$,
designed for the study of particles containing \bquark or \cquark
quarks. The detector includes a high-precision tracking system
consisting of a silicon-strip vertex detector (\velo) surrounding the $pp$
interaction region, a large-area silicon-strip detector located
upstream of a dipole magnet with a bending power of about
$4{\rm\,Tm}$, and three stations of silicon-strip detectors and straw
drift tubes placed downstream.
The magnetic field polarity is regularly flipped to reduce the effect of detection asymmetries. The $pp$ collision data recorded with each of the two magnetic field polarities correspond to approximately half of the data sample.
The combined tracking system provides a momentum measurement with
relative uncertainty that varies from 0.4\% at 5\gevc to 0.6\% at 100\gevc,
and an impact parameter resolution of 20\mum for
tracks with high transverse momentum (\pt). Charged hadrons are identified
using two ring-imaging Cherenkov detectors~\cite{LHCb-DP-2012-003}. Photon, electron and
hadron candidates are identified by a calorimeter system consisting of
scintillating-pad and preshower detectors, an electromagnetic
calorimeter and a hadronic calorimeter. Muons are identified by a
system composed of alternating layers of iron and multiwire
proportional chambers.

Simulated samples are used to determine efficiencies and the probability density functions (PDFs) used in the fits. The $pp$ collisions are generated using \pythia~6.4~\cite{Sjostrand:2006za} with a specific \lhcb
configuration~\cite{LHCb-PROC-2010-056}.  Decays of hadronic particles
are described by \evtgen~\cite{Lange:2001uf}, in which final state
radiation is generated using \photos~\cite{Golonka:2005pn}. The
interaction of the generated particles with the detector and its
response are implemented using the \geant
toolkit~\cite{Allison:2006ve, *Agostinelli:2002hh} as described in
Ref.~\cite{LHCb-PROC-2011-006}.

The trigger~\cite{LHCb-DP-2012-004} consists of a
hardware stage, based on information from the calorimeter and muon
systems, followed by a software stage, which performs a full event
reconstruction.
The candidates used in this analysis are triggered at the hardware stage either directly by one of the particles from the \B candidate decay depositing a transverse energy of at least $3.6\gev$ in the calorimeters, or by other activity in the event (usually associated with the decay products of the other \bquark-hadron decay produced in the $\proton\proton\to\bquark\bquarkbar\PX$ interaction). Inclusion
of the latter category increases the acceptance of signal decays by approximately a factor two.
The software trigger requires a two- or three-particle
secondary vertex with a high scalar sum of the \pt of 
the particles and significant displacement from the primary $pp$ interaction vertices~(PVs). A multivariate algorithm~\cite{BBDT} is used for
  the identification of secondary vertices consistent with the decay
  of a \bquark hadron.
  
Candidate \ButoKsPi and \ButoKsK decays are formed by combining a \decay{\KS}{\pip\pim} candidate with a charged track that is identified as a pion or kaon, respectively. Only tracks in a fiducial volume with small detection asymmetries~\cite{LHCB-PAPER-2011-023} are accepted in the analysis. Pions used to reconstruct the \KS decays are required to have momentum $\ptot>2\gevc$,
$\chisqip>9$, and track segments in the VELO and in the downstream tracking chambers. The \chisqip is defined as the difference in \chisq of a given PV reconstructed with and without the considered particle. The \KS candidates have $\ptot>8\gevc$, $\pt>0.8\gevc$, a good quality vertex fit, a mass within $\pm15\mevcc$ of the known value~\cite{PDG2012}, and are well-separated from all PVs in the event. It is also required that their momentum vectors do not point back to any of the PVs in the event.

Pion and kaon candidate identification is based on the information provided by the RICH detectors \cite{LHCb-DP-2012-003}, combined in the difference in the logarithms of the likelihoods for the kaon and pion hypotheses (\dllkpi). A track is identified as a pion (kaon)
if $\dllkpi \leq 3$ ($\dllkpi > 3$), and $\ptot<110\gevc$, a momentum beyond which there is little separation between pions and kaons.
The efficiencies of these requirements are 95\% and 82\% for signal pions and kaons, respectively. The misidentification probabilities of pions to kaons and kaons to pions are 5\% and 18\%. These figures are determined using a large sample of \decay{\Dstarp}{\Dz(\to\Km\pip)\pip} decays reweighted by the kinematics of the simulated signal decays. Tracks that are consistent with particles leaving hits in the muon detectors are rejected. Pions and kaons are also required to have $\pt>1\gevc$ and $\chisqip>2$. 

The \B candidates are required to have the scalar \pt sum of the \KS and the \pip (or \Kp) candidates that exceeds $4\gevc$, to have $\chisqip<10$ and $\ptot >25\gevc$ and to form a good-quality vertex well separated from all the PVs in the event and displaced from the associated PV by at least $1\mm$. The daughter (\KS or \pip/\Kp) with the larger \pt is required to have an impact parameter above $50 \mum$.
The angle $\theta_{\textrm{dir}}$ between the \B candidate's line of flight and its momentum is required to be less than 32\mrad. 
Background for \KS candidates is further reduced by requiring the \KS decay vertex to be significantly displaced from the reconstructed \B decay vertex along the beam direction ($z$-axis), with $S_z \equiv (z_{\KS}-z_{\B})/\sqrt{\sigma^2_{z,\KS}+\sigma^2_{z,\B}}>2$, where $\sigma^2_{z,\KS}$ and $\sigma^2_{z,\B}$ are the uncertainties on the $z$ positions of the \KS and \B decay vertices $z_{\KS}$ and $z_{\B}$, respectively.

Boosted decision trees~(BDT)~\cite{Breiman} are trained using
the AdaBoost algorithm\cite{AdaBoost} to further separate signal from
background. 
The discriminating variables used are the following: $S_z$; the \chisqip of the \KS and \pip/\Kp candidates; \pt, $\cos(\theta_{\textrm{dir}})$, \chisqvs of the \B candidates defined as the difference in \chisq of fits in which the \Bp decay vertex is constrained to coincide with the PV or not; and the imbalance of \pt, $A_{\pt} \equiv (\pt(\B)-\sum{\pt})/(\pt(\B)+\sum{\pt})$ where the scalar \pt sum is for all the tracks not used to form the \B candidate and which lie in a cone around the \B momentum vector. This cone is defined by a circle of radius 1 unit in the pseudorapidity-azimuthal angle plane, where the azimuthal angle is measured in radians. Combinatorial background tends to be less isolated with smaller \pt imbalance than typical \bquark-hadron decays. The background training samples are taken from the upper \B invariant mass sideband region in data ($5450<m_{\B}<5800\mevcc$), while those of the signal are taken from simulated \ButoKsPi and \ButoKsK decays. Two discriminants are constructed to avoid biasing the background level in the upper \B mass sideband while making maximal use of the
available data for training the BDT. The \KSPip and \KSKp samples are merged to prepare the two BDTs. They are trained using two independent equal-sized subsamples, each corresponding to half of the whole data sample. Both BDT outputs are found to be in agreement with each other in all aspects and each of them is applied to the other sample. 
For each event not used to train the BDTs, one of the two BDT outputs is arbitrarily applied. 
In this way, both BDT discriminants are applied to equal-sized data samples and the number of events used to train the BDTs is maximised without bias of the sideband region and the simulated samples used for the efficiency determination.  
The choice of the requirement on the BDT output (${\cal Q}$) is performed independently for the \KSPi and \KSK samples by evaluating the signal significance $N_{\rm S}/\sqrt{N_{\rm S}+N_{\rm B}}$, where $N_{\rm S}$ ($N_{\rm B}$) denotes the expected number of signal (background) candidates. The predicted effective pollution from mis-identified \ButoKsPi decays in the \ButoKsK signal mass region is taken into account in the calculation of $N_{\rm B}$. The expected signal significance is maximised by applying ${\cal Q}>0.4$ (0.8) for \ButoKsPi (\ButoKsK) decays.

\section{Asymmetries and signal yields}
\label{sec:Yields}

The \CP-summed \ButoKsK and \ButoKsPi yields are measured together with the raw charge asymmetries by means of a simultaneous unbinned extended maximum likelihood fit to the $\B^{\pm}$ candidate mass distributions of the four possible final states (\decay{\B^{\pm}}{\KS\pipm} and \decay{\B^{\pm}}{\KS\Kpm}). Five components contribute to each of the mass distributions. The signal is described by the sum of a Gaussian distribution and a Crystal Ball function (CB) \cite{Skwarnicki:1986xj} with identical peak positions determined in the fit. The CB component models the radiative tail. The other parameters, which are determined from fits of simulated samples, are common for both decay modes. The width of the CB function is, according to the simulation, fixed to be 0.43 times that of the Gaussian distribution, which is left free in the fit.

Due to imperfect particle identification, \ButoKsPi (\ButoKsK) decays can be misidentified as \KSKp (\KSPip) candidates. The corresponding PDFs are empirically modelled with the sum of two CB functions. For the \ButoKsPi (\ButoKsK) decay, the misidentification shape has a significant high (low) mass tail. The parameters of the two CB functions are determined
from the simulation, and then fixed in fits to data.

Partially reconstructed decays, coming mainly from \Bz and \Bp (labelled \B in this section), and \Bs meson decays to open charm and to a lesser extent from three-body charmless \B and \Bs decays, are modelled with two PDFs. These PDFs are identical in the four possible final states. They are modelled by a step function with a threshold mass equal to $m_{\textrm{\B}}-m_{\textrm{\pion}}$ ($m_{\textrm{\Bs}}-m_{\textrm{\pion}}$)~\cite{PDG2012} for \B (\Bs) decays, convolved with a Gaussian distribution of width $20\mevcc$ to account for detector resolution effects. Backgrounds from \Lb decays are found to be negligible. 
The combinatorial background is assumed to have a flat distribution in all categories.

The signal and background yields are varied in the fit, apart from those of the cross-feed contributions, which are constrained using known ratios of selection efficiencies from the simulation and particle identification and misidentification probabilities. The ratio of \ButoKsK (\ButoKsPi) events reconstructed and selected as \KSPip (\KSKp) with respect to \KSKp (\KSPip) are $0.245 \pm 0.018$ ($0.0418 \pm 0.0067$), where the uncertainties are dominated by the finite size of the simulated samples. These numbers appear in Gaussian terms inserted in the fit likelihood function. The charge asymmetries of the backgrounds vary independently in the fit, apart from those of the cross-feed contributions, which are identical to those of the properly reconstructed signal decay.

\begin{figure}[tb]
  \begin{center}
    \includegraphics[width=0.49\linewidth]{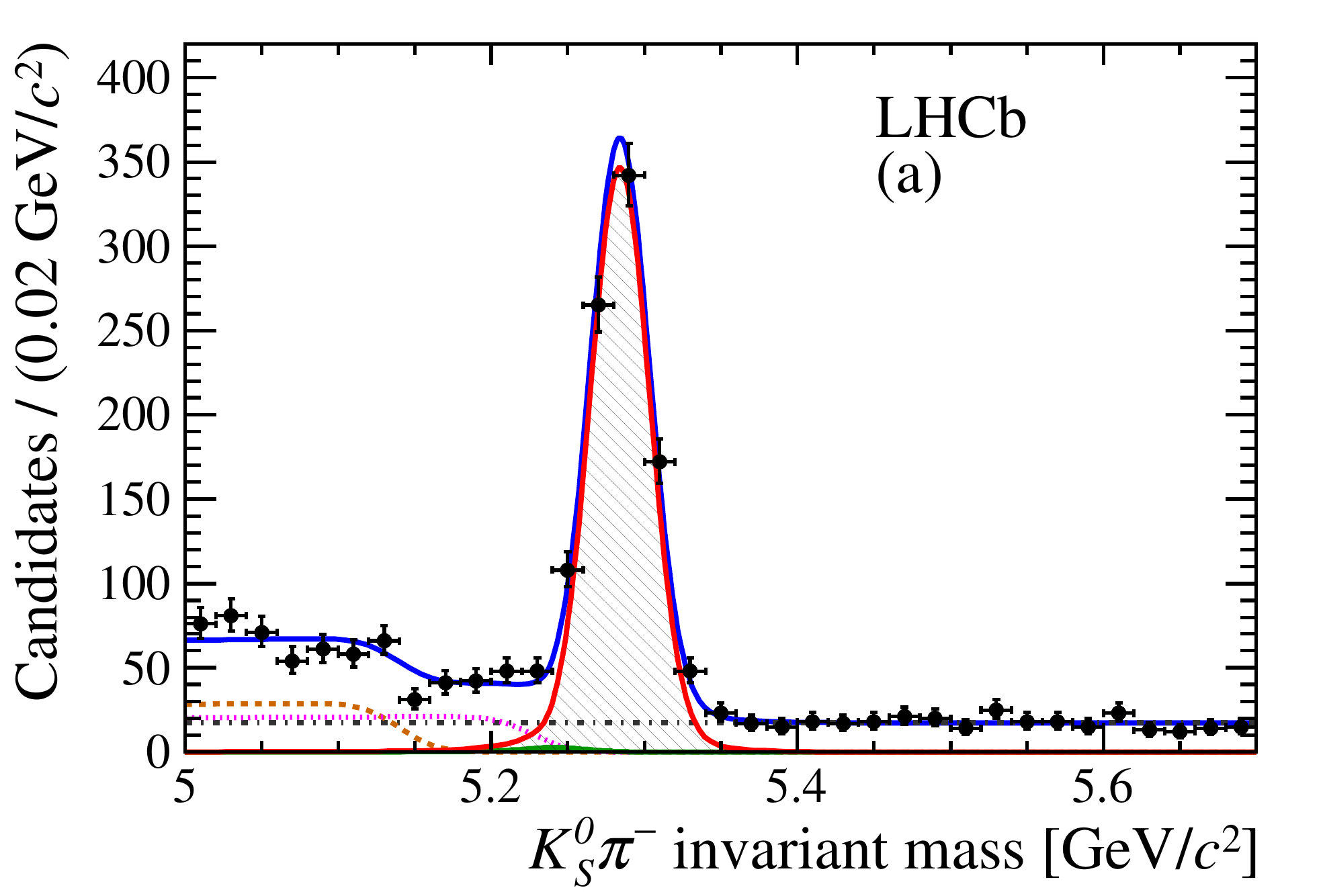}
    \includegraphics[width=0.49\linewidth]{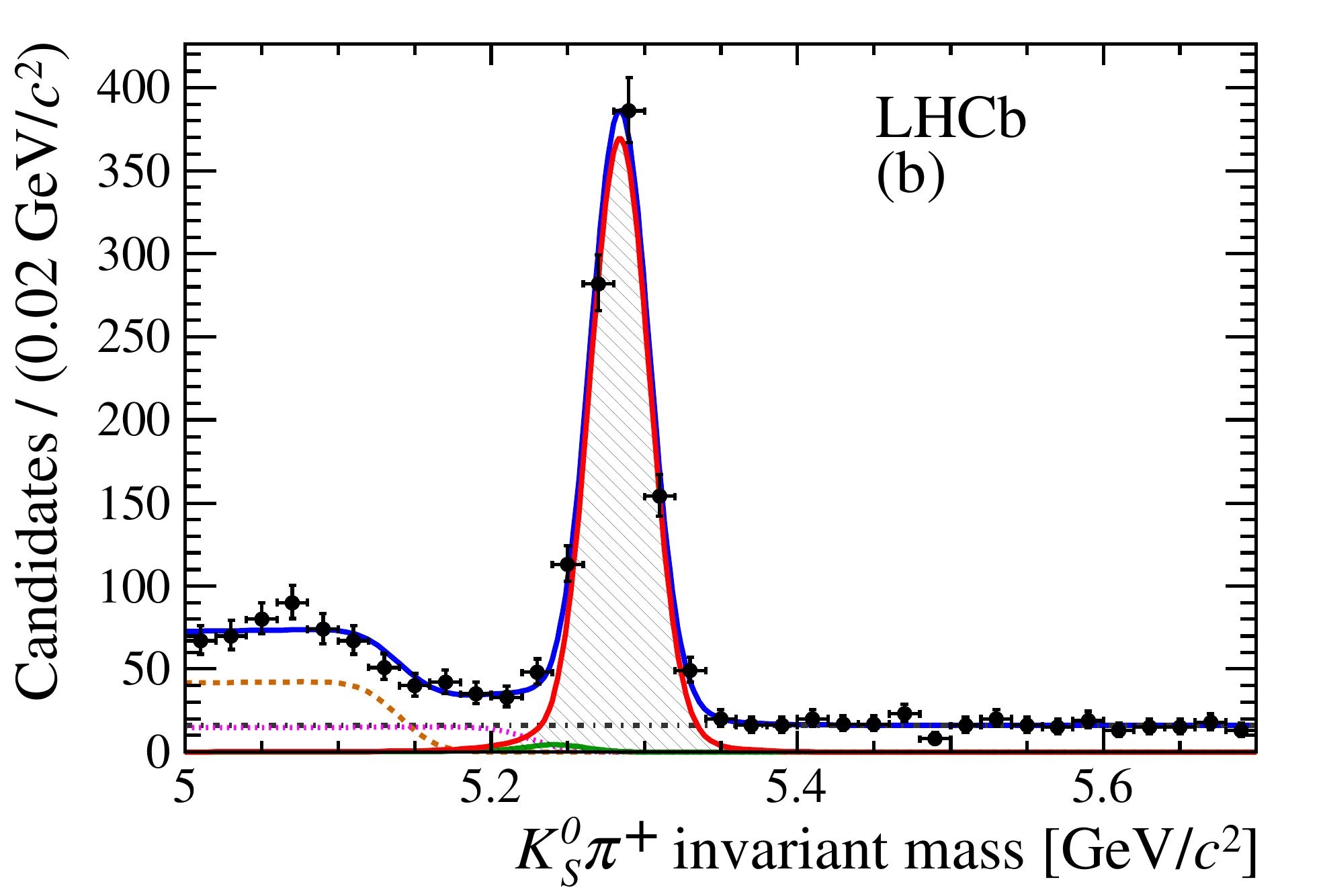}
    \includegraphics[width=0.49\linewidth]{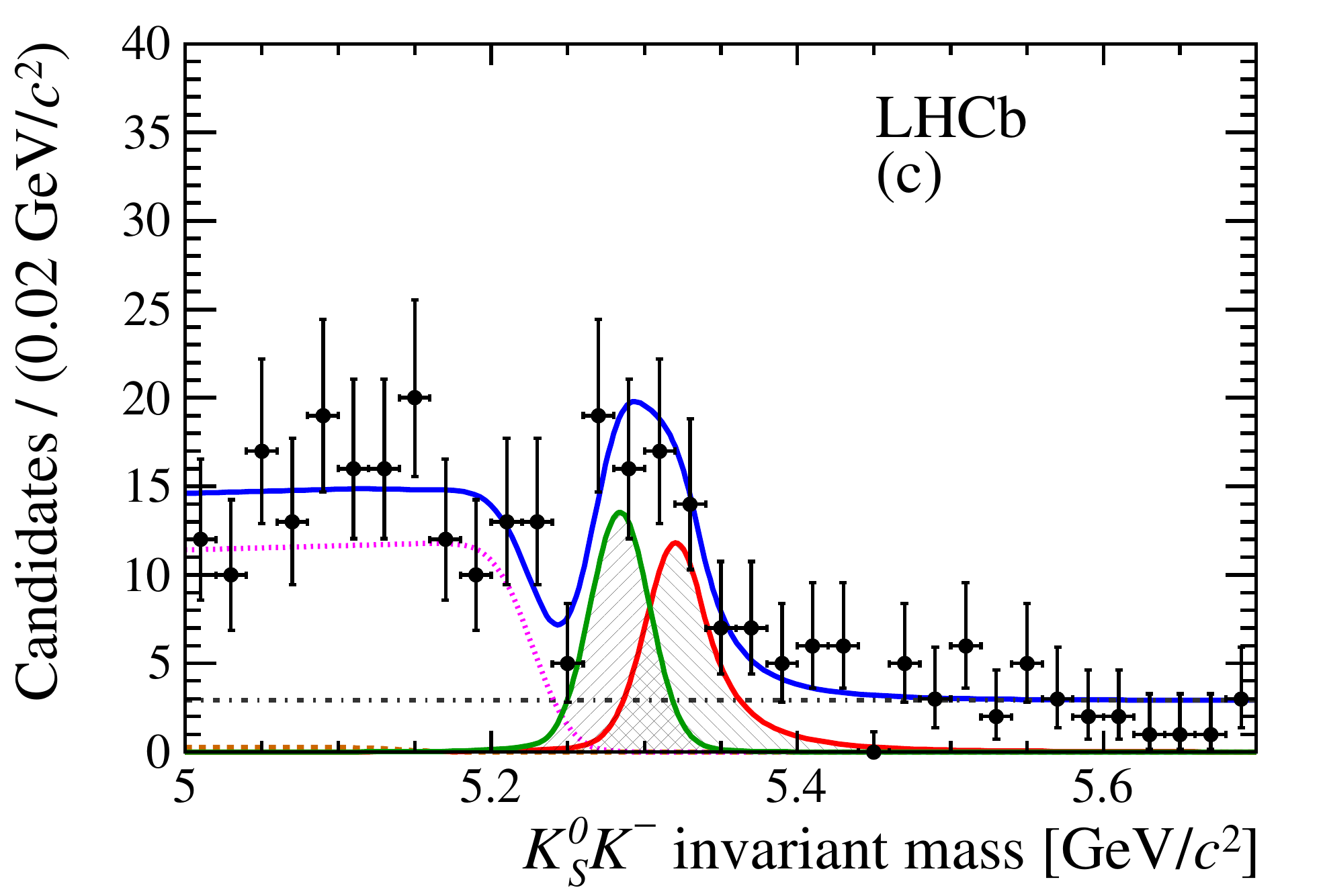}
    \includegraphics[width=0.49\linewidth]{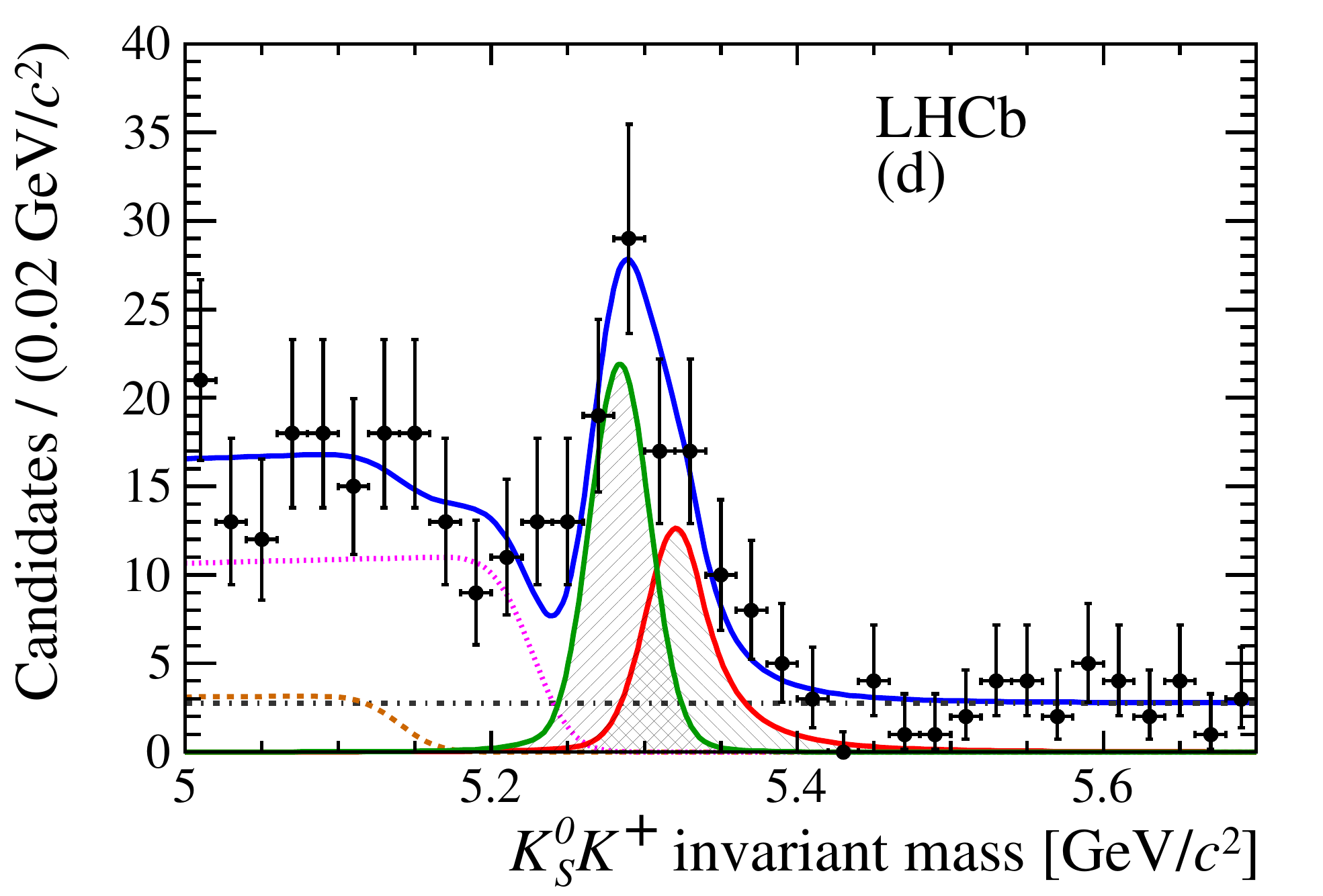}
    \vspace*{-0.75cm}
  \end{center}
  \caption{
    \small 
    Invariant mass distributions of selected (a) \BmtoKSPi, (b) \BptoKSPi, (c) \BmtoKSK and (d) \BptoKSK candidates.
    Data are points with error bars, the \ButoKsPi (\ButoKsK) components are shown as red falling hatched (green rising hatched) curves, combinatorial background is grey dash-dotted, partially reconstructed \Bs (\Bz/\Bu) backgrounds are dotted magenta (dashed orange).
   }
  \label{fits}
\end{figure}

Figure~\ref{fits} shows the four invariant mass distributions along with the projections of the fit. The measured width of the Gaussian distribution used in the signal PDF is found to be approximately 20\% larger than in the simulation, and is included as a systematic uncertainty. The \CP-summed \ButoKsPi and \ButoKsK signal yields are found to be $N(\ButoKsPi)=1804\pm47$ and $N(\ButoKsK)=90\pm13$, with raw \CP asymmetries $\mathcal{A}_{\textrm{raw}}(\ButoKsPi) = -0.032 \pm 0.025$ and $\mathcal{A}_{\textrm{raw}}(\ButoKsK) = -0.23 \pm 0.14$. All background asymmetries are found to be consistent with zero within two standard deviations.  By dividing the sample in terms of data taking periods and magnet polarity, no discrepancies of more than two statistical standard deviations are found in the raw \CP asymmetries.

\section{Corrections and systematic uncertainties}

The ratio of branching fractions is determined as
\begin{eqnarray}
\label{eq:MasterFormulaBR}
\frac{\Br{\ButoKsK}}{\Br{\ButoKsPi}} & = & \frac{N(\ButoKsK)}{N(\ButoKsPi)} \cdot r_{\textrm{sel}} \cdot r_{\textrm{PID}}\textrm{,}
\end{eqnarray}
where the ratio of selection efficiencies is factorised into two terms representing the particle identification,
\begin{eqnarray}
r_{\textrm{PID}} & \equiv & \frac{\eps_{\textrm{PID}}(\ButoKsPi)}{\eps_{\textrm{PID}}(\ButoKsK)},
\end{eqnarray}
and the rest of the selection,
\begin{eqnarray}
r_{\textrm{sel}}  & \equiv  & \frac{\eps_{\textrm{sel}}(\ButoKsPi)}{\eps_{\textrm{sel}}(\ButoKsK)}.
\end{eqnarray}
The raw \CP asymmetries of the \ButoKsPi and \ButoKsK decays are corrected for detection and production asymmetries $\mathcal{A}_{\textrm{det+prod}}$, as well as for a small contribution due to \CP violation in the neutral kaon system ($\mathcal{A}_{\textrm{\KS}}$). The latter is assumed to be the same for both \ButoKsPi and  \ButoKsK decays. At first order, the \ButoKsPi \CP asymmetry can be written as
\begin{eqnarray}
\nonumber
\Acp{\ButoKsPi}  & \approx & \mathcal{A}_{\textrm{raw}}(\ButoKsPi) - \mathcal{A}_{\textrm{det+prod}}(\ButoKsPi) + \mathcal{A}_{\textrm{\KS}}
\end{eqnarray}
and similarly for \ButoKsK, up to a sign flip in front of $\mathcal{A}_{\textrm{\KS}}$.

Selection efficiencies are determined from simulated samples generated at a centre-of-mass energy of 8\tev. The ratio of selection efficiencies is found to be $r_{\textrm{sel}} = 1.111 \pm 0.019$, where the uncertainty is from the limited sample sizes. To first order, effects from imperfect simulation should cancel in the ratio of efficiencies. In order to assign a systematic uncertainty for a potential deviation of the ratio of efficiencies in 7\tev data with respect to 8\tev, the \ButoKsPi and \ButoKsK simulated events are reweighted by a linear function of the \B-meson momentum such that the average \B momentum is 13\% lower, corresponding to the ratio of beam energies. The 0.7\% relative difference between the nominal and reweighted efficiency ratio is assigned as a systematic uncertainty. The distribution of the BDT output for simulated \ButoKsPi events is found to be consistent with the observed distribution of signal candidates in the data using the \emph{sPlot} technique \cite{Pivk:2004ty}, where the discriminating variable is taken to be the \B invariant mass. The total systematic uncertainty related to the selection is 1.8\%. 

The determination of the trigger efficiencies is subject to variations in the data-taking conditions and, in particular, to the ageing of the calorimeter system. These effects are mitigated by regular changes in the gain of the calorimeter system. A large sample of \decay{\Dstarp}{\Dz(\to\Km\pip)\pip} decays is used to measure the trigger efficiency in bins of \pt for pions and kaons from signal decays. These trigger efficiencies are averaged using the \pt distributions obtained from simulation. The hardware stage trigger efficiencies obtained by this procedure are in agreement with those obtained in the simulation within 1.1\%, which is assigned as systematic uncertainty on the ratio of branching fractions. The same procedure is also applied to \Bp and \Bm decays separately, and results in 0.5\% systematic uncertainty on the determination of the \CP asymmetries.

Particle identification efficiencies are determined using a large sample of \decay{\Dstarp}{\Dz(\to\Km\pip)\pip} decays. The kaons and pions from this calibration sample are reweighted in 18 bins of momentum and 4 bins of pseudorapidity, according to the distribution of signal kaons and pions from simulated \ButoKsK and \ButoKsPi decays. The ratio of efficiencies is $r_{\textrm{PID}} = 1.154 \pm 0.025$, where the uncertainty is given by the limited size of the simulated samples. The systematic uncertainty associated with the binning scheme is determined by computing the deviation of the average efficiency calculated using the nominal binning from that obtained with a single bin in each kinematic variable. A variation of $0.7\%$ (1.3\%) is observed for pions (kaons). A systematic uncertainty of $0.5\%$ is assigned due to variations of the efficiencies, determined by comparing results obtained with the 2011 and 2012 calibration samples. All these contributions are added in quadrature to obtain 2.7\% relative systematic uncertainty on the particle identification efficiencies. Charge asymmetries due to the PID requirements are found to be negligible.

Uncertainties due to the modelling of the reconstructed invariant mass distributions are assigned by generating and fitting pseudo-experiments. Parameters of the signal and cross-feed distributions are varied according to results of independent fits to the \ButoKsK and \ButoKsPi simulated samples. 
 The relative uncertainty on the ratio of yields from mis-modelling of the signal (cross-feed) is 2.4\% (2.7\%) mostly affecting the small \ButoKsK yield. The width of the Gaussian resolution function used to model the partially reconstructed backgrounds is increased by 20\%, while the other fixed parameters of the partially reconstructed and combinatorial backgrounds are left free in the fit, in turn, to obtain a relative uncertainty of $3.3\%$. The total contribution of the fit model to the systematic uncertainty is $4.9\%$. Their contribution to the systematic uncertainties on the \CP asymmetries is found to be negligible.

Detection and production asymmetries are measured using approximately one million \ButoJpsiK decays collected in 2011 and 2012. Using a kinematic and topological selection similar to that employed in this analysis, a high purity sample is obtained. The raw \CP asymmetry is measured to be $\mathcal{A}(\ButoJpsiK) = (-1.4 \pm 0.1)\%$ within $20\mevcc$ of the \Bu meson mass. The same result is obtained by fitting the reconstructed invariant mass with a similar model to that used for the \ButoKsPi and \ButoKsK fits.  
This asymmetry is consistent between bins of momentum and pseudorapidity within 0.5\%, which is assigned as the corresponding uncertainty. The \CP asymmetry in \ButoJpsiK decays is $\Acp{\ButoJpsiK} = (+0.5 \pm 0.3)\%$, where the value is the weighted average of the values from Refs.~\cite{PDG2012} and \cite{Abazov:2013sqa}. This leads to a correction of $\mathcal{A}_{\textrm{det+prod}}(\ButoKsK) = (-1.9 \pm 0.6)\%$. 
The combined production and detection asymmetry for \ButoKsPi decays is expressed as $\mathcal{A}_{\textrm{det+prod}}(\ButoKsPi) = \mathcal{A}_{\textrm{det+prod}}(\ButoKsK) + \mathcal{A}_{\textrm{\kpi}}$, where the kaon-pion detection asymmetry is $\mathcal{A}_{\textrm{\kpi}} \approx \mathcal{A}_{\textrm{\kaon}}-\mathcal{A}_{\textrm{\pion}} = (1.0 \pm 0.5) \%$ \cite{LHCb-PAPER-2013-018}. 
  The assigned uncertainty takes into account a potential dependence of the difference of asymmetries as a function of the kinematics of the tracks. The total correction to \Acp{\ButoKsPi} is $\mathcal{A}_{\textrm{det+prod}}(\ButoKsPi) = (-0.9 \pm 0.8)\%$.

Potential effects from \CP violation in the neutral kaon system, either directly via \CP violation in the neutral kaon system \cite{Grossman:2011zk} or via regeneration of a \KS component through interactions of a \KL state with material in the detector \cite{Ko:2010mk}, are also considered. The former is estimated \cite{LHCB-PAPER-2012-026} by fitting the background subtracted \cite{Pivk:2004ty} decay time distribution of the observed \ButoKsPi decays and contributes 0.1\% to the observed asymmetry. 
The systematic uncertainty on this small effect is chosen to have the same magnitude as the correction itself. The latter has been studied \cite{LHCB-PAPER-2012-052} and is small for decays in the \lhcb acceptance and thus no correction is applied. The systematic uncertainty assigned for this assumption is estimated by using the method outlined in Ref.~\cite{Ko:2010mk}. Since the \KS decays reconstructed in this analysis are concentrated at low lifetimes, the two effects are of similar sizes and have the same sign. Thus an additional systematic uncertainty equal to the size of the correction applied for \CP violation in the neutral kaon system and 100\% correlated with it, is assigned. It results in $\mathcal{A}_{\textrm{\KS}} = (0.1 \pm 0.2)\%$. 
A summary of the sources of systematic uncertainty and corrections to the \CP asymmetries are given in Table~\ref{tab:systematics}. Total systematic uncertainties are calculated as the sum in quadrature of the individual contributions.

\begin{table}[t]
  \caption{
    \small Corrections (above double line) and systematic uncertainties (below double line). The relative uncertainties on the ratio of branching fractions are given in the first column. The absolute corrections and related uncertainties on the \CP asymmetries are given in the next two columns. The last column gathers the relative systematic uncertainties contributing to $r_{\Bcp}$. All values are given as percentages.
    }
\begin{center}\begin{tabular}{l|cccc}
    \hline
    \rule{0pt}{2.5ex} Source  & $\mathcal{B}$ ratio & \Acp{\ButoKsPi} & \Acp{\ButoKsK}  & \Bc \\ 
      \hline
    $\mathcal{A}_{\textrm{det+prod}}$  & - & $-0.9$ & $-1.9$ & - \\
    $\mathcal{A}_{\textrm{\KS}}$  & - & 0.1 & 0.1 & - \\
    \hline
    \hline
    Selection & 1.8 & - & - & 6.1\\
    Trigger   & 1.1 & 0.5 & 0.5 & 1.1\\
    Particle identification & 2.7 & - & - & 3.6\\
    Fit model & 4.9 & - & - & 2.0\\
    $\mathcal{A}_{\textrm{det+prod}}$ & - & 0.8 & 0.6 & -\\
    $\mathcal{A}_{\textrm{\KS}}$ & - & 0.2 & 0.2 & -\\
    \hline
    Total syst. uncertainty & 6.0 & 1.0 & 0.8 & 7.4\\ 
    \hline
  \end{tabular}\end{center}
\label{tab:systematics}
\end{table}

\section{Search for \boldmath{\BctoKsK} decays}
\label{sec:Bcsearch}

\begin{figure}[tbp]
  \begin{center}
    \includegraphics[width=0.49\linewidth]{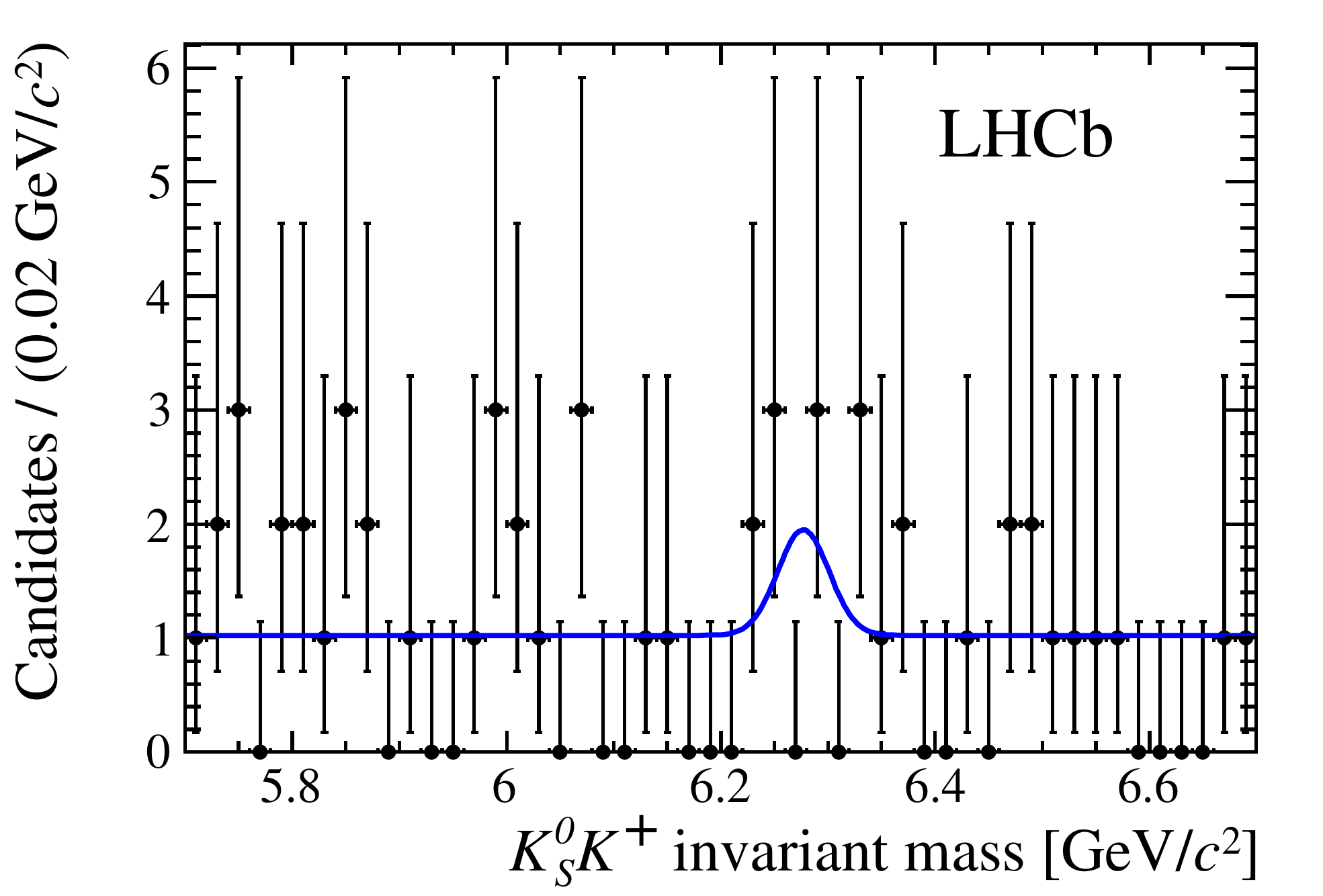}
    \includegraphics[width=0.49\linewidth]{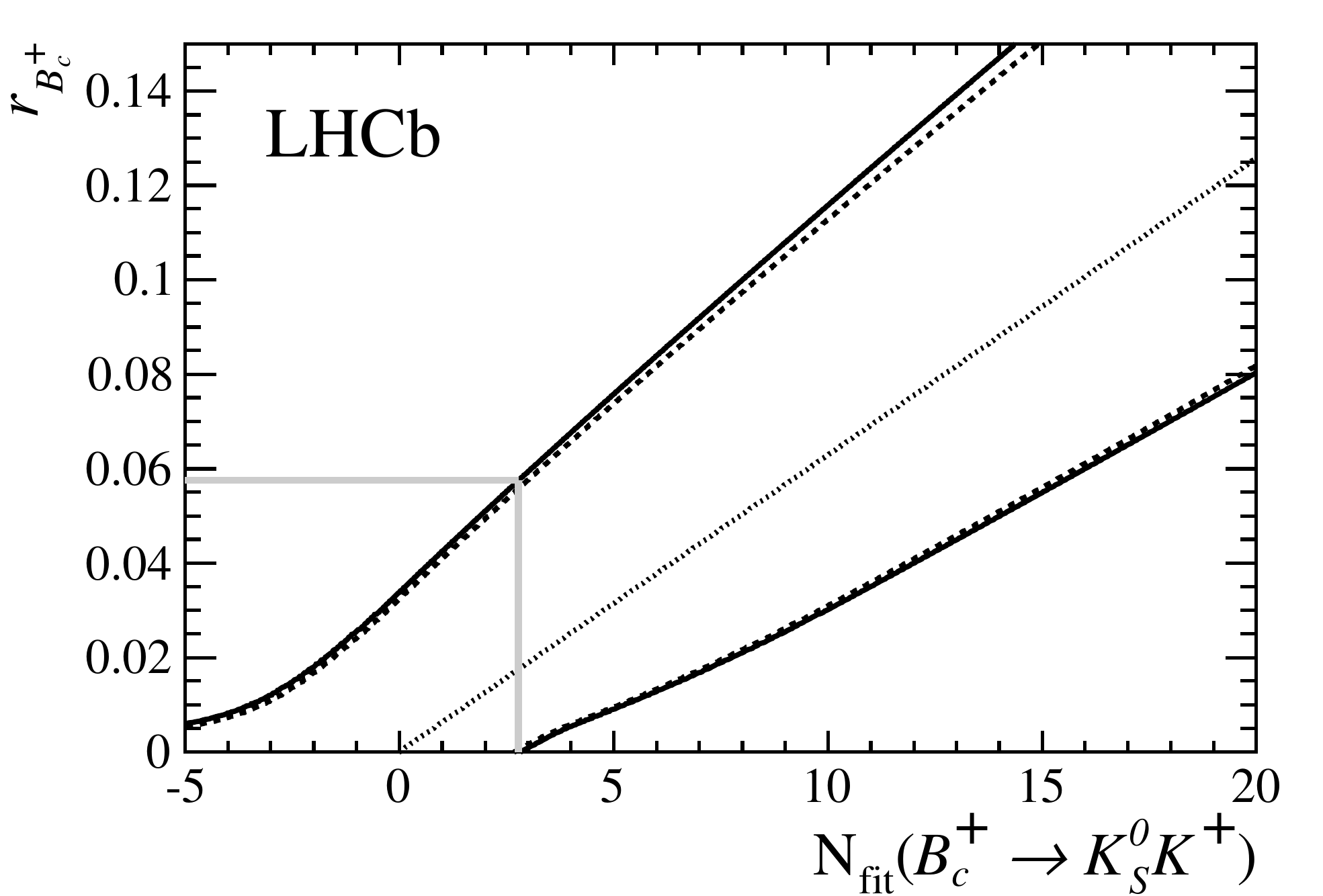}
      \vspace*{-0.75cm}
  \end{center}
  \caption{
    \small
     (Left) Invariant mass distribution of selected \BctoKsK candidates.
     Data are points with error bars and the curve represents the fitted function.   
    (Right) The number of events and the corresponding value of $r_{\Bc}$.
The central value (dotted line) and the upper and lower 90\% statistical
confidence region bands are obtained using the Feldman and Cousins approach \cite{Feldman:1997qc} (dashed lines).
The solid lines includes systematic uncertainties.
The gray outline of the box shows the obtained upper limit of  $r_{\Bc}$ for the observed number of 2.8 events.
   }
  \label{fig2}
\end{figure}

An exploratory search for \BctoKsK decays is performed with the data sample collected in 2011, corresponding to an integrated luminosity of 1\invfb. 
 The same selection as for the \ButoKsK decays is used, only adding a proton veto $\dllpk<10$ to the \Kp daughter, which is more than $99\%$ efficient. This is implemented to reduce a significant background from baryons in the invariant mass region considered for this search. The ratios of selection and particle identification efficiencies are $r_{\rm sel} = 0.306 \pm 0.012$ and $r_{\rm PID} = 0.819 \pm 0.027$, where the uncertainties are from the limited size of the simulated samples. The related systematic uncertainties are estimated in a similar way as for the  measurement of $\Br{\ButoKsK}/\Br{\ButoKsPi}$. The \ButoKsPi yield is also evaluated with the 2011 data only. The \Bcp signal yield is determined by fitting a single Gaussian distribution with the mean fixed to the \Bc mass \cite{PDG2012} and the width fixed to $1.2$ times the value obtained from simulation to take into account the worse resolution in data. The combinatorial background is assumed to be flat. The invariant mass distribution and the superimposed fit are presented in Fig.~\ref{fig2} (left).  Pseudo-experiments are used to evaluate the biases in the fit procedure and the systematic uncertainties are evaluated by assuming that the combinatorial background has an exponential slope. A similar procedure is used to take into account an uncertainty related to the assumed width of the signal distribution. The $20\%$ correction applied to match the observed resolution in data, is assumed to estimate this uncertainty.
 
 The Feldman and Cousins approach \cite{Feldman:1997qc} is used to build 90\% confidence region bands that relate the true value of $r_{\Bc} = (f_c\cdot\Br{\BctoKsK})/(f_u\cdot\Br{\ButoKsPi})$ to the measured number of signal events, and where $f_c$ and $f_u$ are the hadronisation fraction of a \bquark into a \Bcp and a \Bp meson, respectively. All of the systematic uncertainties are included in the construction of the confidence region bands by inflating the width of the Gaussian functions used to build the ranking variable of the Feldman and Cousins procedure. The result is shown in Fig.~\ref{fig2} (right) and gives the upper limit
\begin{equation}
\nonumber
r_{\Bc} \equiv \frac{f_c}{f_u}\cdot\frac{\Br{\BctoKsK}}{\Br{\ButoKsPi}}   <  5.8\times10^{-2} \textrm{ at 90\% confidence level.}
\end{equation}
This is the first upper limit on a \Bcp meson decay into two light quarks.

\section{Results and summary}
\label{sec:Results}
The decays \ButoKsK and \ButoKsPi have been studied using a data sample corresponding to an integrated luminosity of 3\invfb, collected in 2011 and 2012 by the \lhcb detector and the ratio of branching fractions and \CP asymmetries are found to be 
\begin{eqnarray}
\nonumber
\frac{\Br{\ButoKsK}}{\Br{\ButoKsPi}} &=& \phantom{-}0.064 \pm 0.009\textrm{ (stat.)} \pm 0.004\textrm{ (syst.)},\\
\nonumber
\Acp{\ButoKsPi} &=& -0.022 \pm 0.025\textrm{ (stat.)} \pm 0.010\textrm{ (syst.)},
\end{eqnarray}
and
\begin{eqnarray}
\nonumber
\Acp{\ButoKsK}  &=& -0.21 \pm 0.14 \textrm{ (stat.)} \pm 0.01 \textrm{ (syst.)}.
\end{eqnarray}
These results are compatible with previous determinations \cite{Aubert:2006gm,Duh:2012ie}. The measurements of  \Acp{\ButoKsK} and $\Br{\ButoKsK}/\Br{\ButoKsPi}$ are the best single determinations to date. A search for \BctoKsK decays is also performed with a data sample corresponding to an integrated luminosity of 1\invfb. The upper limit 
\begin{eqnarray}
\nonumber
\frac{f_c}{f_u}\cdot\frac{\Br{\BctoKsK}}{\Br{\ButoKsPi}}   <  5.8\times10^{-2} \textrm{ at 90\% confidence level}
\end{eqnarray}
is obtained.
Assuming $f_c \simeq 0.001$~\cite{PhysRevD.80.114031}, $f_u = 0.33$~\cite{PDG2012,LHCb-PAPER-2011-018,LHCb-PAPER-2012-037}, and $\Br{\ButoKzPi} = (23.97\pm0.53\textrm{ (stat.)} \pm 0.71\textrm{ (syst.)})\cdot 10^{-6}$ \cite{Duh:2012ie}, an upper limit $\Br{\BctoKzK} < 4.6\times 10^{-4} \textrm{ at 90\% confidence level}$ is obtained. This is about two to four orders of magnitude higher than theoretical predictions, which range from $10^{-8}$ to $10^{-6}$~\cite{PhysRevD.80.114031}. With the large data samples already collected by the \lhcb experiment, other two-body \Bcp decay modes to light quarks such as \BctoKstK and \BctoPhiK may be searched for.

\section*{Acknowledgements}

\noindent We express our gratitude to our colleagues in the CERN
accelerator departments for the excellent performance of the LHC. We
thank the technical and administrative staff at the LHCb
institutes. We acknowledge support from CERN and from the national
agencies: CAPES, CNPq, FAPERJ and FINEP (Brazil); NSFC (China);
CNRS/IN2P3 and Region Auvergne (France); BMBF, DFG, HGF and MPG
(Germany); SFI (Ireland); INFN (Italy); FOM and NWO (The Netherlands);
SCSR (Poland); MEN/IFA (Romania); MinES, Rosatom, RFBR and NRC
``Kurchatov Institute'' (Russia); MinECo, XuntaGal and GENCAT (Spain);
SNSF and SER (Switzerland); NAS Ukraine (Ukraine); STFC (United
Kingdom); NSF (USA). We also acknowledge the support received from the
ERC under FP7. The Tier1 computing centres are supported by IN2P3
(France), KIT and BMBF (Germany), INFN (Italy), NWO and SURF (The
Netherlands), PIC (Spain), GridPP (United Kingdom). We are thankful
for the computing resources put at our disposal by Yandex LLC
(Russia), as well as to the communities behind the multiple open
source software packages that we depend on.

\addcontentsline{toc}{section}{References}
\setboolean{inbibliography}{true}
\bibliographystyle{LHCb}
\bibliography{main,LHCb-PAPER,LHCb-CONF,LHCb-DP}

\end{document}